\documentstyle[tighten,preprint,aps]{revtex}
\begin{document}
\title{Measurements at low energies of the polarization-transfer
  coefficient $K_{y}^{y'}$ for the reaction
  $^{3}$H$(\vec{p},\vec{n})^{3}$He at $0^{\circ}$}
\author{W.S.~Wilburn,$^{(1)}$\thanks{Present address:
    Los Alamos National Laboratory, Los Alamos, NM 87545}
  C.R.~Gould,$^{(2)}$
  G.M.~Hale,$^{(3)}$
  P.R.~Huffman,$^{(1)}$\thanks{Present address:
    Department of Physics, Harvard University, Cambridge, MA 02138}
  C.D.~Keith,$^{(2)}$\thanks{Present address:
    Thomas Jefferson National Accelerator Facility, Newport News, VA 23606}
  N.R.~Roberson,$^{(1)}$ and
  W.~Tornow$^{(1)}$}
\address{$^{(1)}$ Physics Department, Duke University, Durham, NC
  27708 \\
  and Triangle Universities Nuclear Laboratory, Durham, NC 27708 \\
  $^{(2)}$ Physics Department, North Carolina State University,
  Raleigh, NC 27695 \\
  and Triangle Universities Nuclear Laboratory, Durham, NC 27708 \\
  $^{(3)}$ Los Alamos National Laboratory, Los Alamos, NM 87545}
\date{\today}
\maketitle

\begin{abstract}
  Measurements of the transverse polarization coefficient $K_{y}^{y'}$
  for the reaction $^{3}$H$(\vec{p},\vec{n})^{3}$He are reported for
  outgoing neutron energies of 1.94, 5.21, and 5.81~MeV\@. This
  reaction is important both as a source of polarized neutrons for
  nuclear physics experiments, and as a test of theoretical
  descriptions of the nuclear four-body system. Comparison is made to
  previous measurements, confirming the
  $^{3}$H$(\vec{p},\vec{n})^{3}$He reaction can be used as a polarized
  neutron source with the polarization known to an accuracy of
  approximately 5\%. Comparison to \emph{R}-matrix theory suggests
  that the sign of the $^{3}\!F_{3}$ phase-shift parameter is
  incorrect. Changing the sign of this parameter dramatically improves
  the agreement between theory and experiment.
\end{abstract}

\section{Introduction}
  \label{sec:int}
  
  A polarized neutron beam is an important tool in studying the
  spin-dependent aspects of the nuclear interaction.  Intense beams of
  polarized neutrons can be produced from reactions involving
  polarized charged-particle beams, provided the reaction has a
  non-zero polarization transfer coefficient. One reaction which is
  often used at few-MeV energies is the
  $^{3}$H$(\vec{p},\vec{n})^{3}$He polarization-transfer reaction.
  This reaction was chosen as a polarized neutron source for recent
  experiments at Triangle Universities Nuclear Laboratory (TUNL)
  \cite{Wil93b,Kei96}. As the transverse polarization transfer
  coefficient $K_{y}^{y'}$ for protons on tritium was measured more
  than 20~years ago \cite{Don71b}, we felt it important to confirm the
  values at energies of interest. In addition, despite advances in the
  \emph{R}-matrix treatment of the four-body system, there is no
  recent comparison with polarization transfer data.
  
  The polarization-transfer measurement is made using a transversely
  polarized proton beam incident on a tritiated target. Two
  polarimeters are used, one for the proton beam, and one for the
  neutron beam. The transverse polarization transfer coefficient
  $K_{y}^{y'}=P_{n}/P_{p}$ is determined by the ratio of the neutron
  polarization, $P_{n}$, to the proton polarization, $P_{p}$.
  
  The measurements were made at neutron energies of 1.94, 5.21, and
  5.81~MeV\@. The results are in excellent agreement with the previous
  data, and confirm that the $^{3}$H$(\vec{p},\vec{n})^{3}$He reaction
  can be used as a polarized neutron source to an accuracy of
  approximately 5\%.
  
  In comparing the data to \emph{R}-matrix predictions, we find a
  discrepancy which we attribute to the sign of the $^{3}\!F_{3}$
  phase shift parameter. Changing the sign of this parameter
  dramatically improves the agreement with experiment, while having
  little effect on other observables.

\section{Experimental Apparatus}
  \label{sec:exp}
  
  The polarized proton beam is produced by the TUNL atomic beam
  polarized ion source (ABPIS) \cite{Cle95b} and accelerated by a
  tandem Van de Graaff. The proton polarization is set by a Wien
  filter spin precessor to be vertical, along the $y$~direction.  The
  neutron production target is a tritiated-titanium foil
  ($3\times10^{10}$~Bq/cm$^{2}$) on an isotopically pure $^{58}$Ni
  backing. This backing material is chosen as the 9.44~MeV threshold
  for the $^{58}$Ni$(p,n)^{58}$Cu reaction greatly reduces the
  background neutron level.
  
  Slit feedback steering is used to control the position of the beam
  after acceleration, keeping the beam centered on the entrance to the
  proton polarimeter. In addition, a beam-profile monitor is used to
  determine the beam centroid 2~m from the neutron production target.
  A computer-controlled feedback system uses this information to
  center the beam in the beam pipe. This system, in combination with a
  slit feedback loop, keeps the beam centered on the neutron
  production foil. The tritiated-titanium foil is isolated from the
  beam line by a 2.5~$\mu$m havar foil.  The intervening space is
  filled with helium gas at a pressure of 0.1~Mpa.

\section{The proton polarimeter}
  \label{sec:ppl}
    
    The proton polarimeter consists of a carbon foil and two solid
    state charged-particle detectors contained within a small
    scattering chamber as shown in Fig.~\ref{fig:ppol}. The aluminum
    chamber is made of a cylindrical body with two arms acting as
    particle flight paths to minimize the internal volume and thus
    reduce the need for vacuum pumping. The detectors are mounted at
    the ends of the arms with three sets of tantalum collimators
    defining the scattered beams with an angular acceptance of
    $\pm3.46^{\circ}$. The carbon foil is 22~mm in diameter and
    5~$\mu$g/cm$^{2}$ thick and is mounted on a plunger so that it can
    be removed from the beam when not in use.  Slits in front of the
    polarimeter define the beam to be $12.7\times12.7$~mm. The slit
    currents are fed back to a steering magnet which keeps the beam
    centered on the entrance of the polarimeter. The detectors and all
    but the first set of collimators are electrically isolated to
    reduce noise pickup.  The detectors are silicon charged-particle
    detectors with 300~mm$^{2}$ active areas and depletion depths of
    1000~$\mu$m.  The angle of the arms is fixed, for simplicity, at
    $\pm40^{\circ}$. This angle is chosen to provide a relatively
    large analyzing power over a range of energies for protons
    elastically scattered from carbon. At the lowest energy of this
    experiment ($E_{p}=3.02$~MeV), however, the analyzing power is
    small. In this case, the proton beam polarization is measured at a
    higher energy before and after the measurement.
  
    The detector signals are amplified and sent to single-channel
    analyzers (SCA). The outputs of the SCA's are counted by scalers
    and stored in the computer. Three well-separated peaks are
    observed in the pulse height spectra with negligible background.
    In order of increasing energy of the detected particle they are
    elastic scattering from hydrogen H$(\vec{p},p)$H, inelastic
    scattering from carbon $^{12}$C$(\vec{p},p_{1})^{12}$C, and
    elastic scattering from carbon $^{12}$C$(\vec{p},p_{0})^{12}$C. In
    the case of elastic scattering from hydrogen, it is the recoiling
    proton which is detected. The SCA windows are set around the
    $^{12}$C elastic scattering peak.

\section{The neutron polarimeter}
  \label{sec:npl}
  
  The neutron polarimeter consists of a $^{4}$He target and neutron
  detector pairs placed symmetrically about the beam direction as
  shown in Fig.~\ref{fig:npol}. Since the analyzing power of neutrons
  scattered elastically from helium can be calculated from the
  \emph{n}-$^{4}$He phase shifts with an accuracy of approximately
  $\pm0.01$ in the energy range of interest, the neutron beam
  polarization can be extracted from the measured left-right
  asymmetry. The polarimeter is located behind a polyethylene
  collimator which defines the neutron beam. The distance from the
  neutron source to the center of the polarimeter target is 1.79~m.
  
  The target consists of helium--xenon gas mixture at a pressure of
  10~MPa in a steel cylinder with 1~mm walls \cite{Tor74}. The active
  volume of the cell has a diameter of 44.6~mm and a height of
  158.2~mm. Upstream neutron collimation defines a beam spot at the
  center of the polarimeter that is 31.5~mm in the horizontal
  direction by 94.5~mm in the vertical direction. The gas cell, shown
  in Fig.~\ref{fig:hecl}, has windows at both ends to allow
  scintillation light from the recoiling $^{4}$He to be detected in
  photomultiplier tubes. Approximately 5\% xenon, by partial pressure,
  is added to the helium to increase the light output. This
  arrangement makes it an active target and allows the measurement of
  both timing coincidences and $^{4}$He recoil energy, thus reducing
  the background considerably. The inner walls of the gas cell are
  coated with 0.8~mm of magnesium oxide as a reflector. The windows
  are opaque to the majority of the scintillation light, whose
  wavelength is centered in the extreme ultraviolet (XUV) spectral
  region. Therefore 120~$\mu$g/cm$^{2}$ of \emph{p}-quaterphenyl is
  evaporated on top of the reflector and acts as a wavelength shifter.
  The photomultiplier tubes are Hamamatsu type~H1161 having a diameter
  of 51~mm. Care is taken while filling the gas cell to eliminate
  contaminants such as nitrogen and oxygen, which greatly reduce the
  performance of the cell by absorbing the light of scintillation.
  
  The neutron detectors are placed symmetrically about $0^{\circ}$ at
  angles where the product of the square of the analyzing power and
  the cross section is a maximum for $^{4}$He$(\vec{n},n)^{4}$He.
  There are two such angle pairs, one forward and one backward, and
  two detector pairs are used.  At the lowest energy measured,
  however, one maximum is too low in magnitude to be useful and only
  one detector pair is used.  The detectors are organic liquid
  scintillators (NE213) coupled to photomultiplier tubes through light
  guides. The light guides match the geometry of the
  $45\times158\times76$~mm thick rectangular scintillator cells to the
  51~mm diameter phototubes. The photomultiplier tubes are Hamamatsu
  type~H1161.  The neutron detectors are mounted to a detector ring
  having $1^{\circ}$ graduations. The entire apparatus is aligned
  optically with the neutron beam collimator.
  
  Monte Carlo techniques are used to calculate the effective analyzing
  power of the neutron polarimeter \cite{Tor75}. These calculations
  make corrections for the finite geometry of the target cell and
  neutron detectors, and for double-scattering events. Double
  scattering which includes neutrons scattering from materials in the
  target cell other than helium must be considered. The
  double-scattering events which are considered are He--He, He--Xe,
  Xe--He, He--Fe, and Fe--He. Since only events which include the
  detection of $^{4}$He recoil are recorded, single scattering events
  from elements other than helium need not be considered. Cross
  sections and analyzing powers are obtained from phase-shift data
  sets. The phase shifts for \emph{n}-$^{4}$He come from experimental
  data \cite{Sta72} as do the phase shifts for \emph{n}-Xe and
  \emph{n}-Fe at the higher energies. At the lowest energy measured,
  1.94~MeV, experimental cross section data for xenon are used with
  analyzing powers assumed to be zero.
  
  For an event to be considered valid, three criteria must be
  satisfied. First, there must be a coincidence between the top and
  bottom photomultiplier tubes of the target cell, reducing the noise
  from the tubes. Second, the signal from the neutron detector must
  meet pulse-shape discrimination requirements, eliminating signals
  from gammas. Third, there must be a coincidence between the target
  cell signal and one of the neutron detector signals, in order to
  reject neutron background events. The anode signals from the target
  cell photomultiplier tubes are used to form the coincidences, while
  the dynode signals provide the $^{4}$He recoil energy information.
  Summing the dynode signals improves the energy resolution of the
  detector as only part of the light of scintillation is deposited in
  each photomultiplier tube. The gains of the two tubes are matched
  using a $^{137}$Cs source. Since the $^{4}$He recoil energy is quite
  different for the forward and backward angles, the summed dynode
  signal is split into two signals which are amplified and delayed by
  different amounts before being recombined. This process creates two
  pulses for each $^{4}$He recoil, one large in amplitude and one
  small. The large amplitude pulse is gated through to the data
  acquisition system only when an event occurs in one of the backward
  angle detectors. Conversely, the small amplitude pulse is gated
  through only when an event occurs in one of the forward angle
  detectors. This arrangement allows a single ADC to digitize the
  $^{4}$He recoil energy for both detector pairs. Neutron
  time-of-flight from the target cell to the neutron detectors is
  determined using Ortec~467 time-to-amplitude converters (TAC)\@. The
  start signal for each TAC comes from the timing signal of the
  corresponding neutron detector after the coincidence with the target
  cell anode coincidence signal. The coincidences are formed such that
  the timing signal from the neutron detector determines the timing of
  the TAC start signal.  The stop signal comes from a delayed version
  of the target cell anode signal. This delay is adjusted to give a
  suitable range for the neutron time-of-flight signals (typically
  100--200~ns).  Because of the coincidence requirements, the count
  rates are low, typically a few per second.

\section{Data acquisition}
  \label{sec:daq}
  
  Data from the measurements are collected by CAMAC modules controlled
  by a MBD-11 Multiple Branch Driver \cite{Rob81}. The MBD is
  connected to a Digital Electronics Corporation VAXStation~3200 which
  supervises the data acquisition, stores the data, and performs
  online analysis. In addition, this computer performs part of the
  feedback steering of the proton beam. The data acquisition software
  runs under the TUNL XSYS data acquisition and analysis system
  \cite{Gou81}.
  
  For the polarization measurements it is necessary to measure
  count-rate asymmetries to an accuracy of order $1\times10^{-3}$.
  Instrumental asymmetries due to the data acquisition system must
  therefore be of order $1\times10^{-4}$ or less. To achieve this
  level of stability, fast spin-reversal is employed. The spin of the
  neutron beam is reversed at a rate of 10~Hz to reduce the effects of
  detector gain drifts and other instrumental shifts. Precision timing
  is used to insure that the same amount of time is spent counting in
  each spin state.
  
  The fast spin-reversal is controlled by a NIM Spin State Control
  (SSC) module. The module produces an eight-step spin sequence:
  $+--+-++-$ which cancels the effects of detector drifts to second
  order in time \cite{Rob93}. All scaler data and ADC gates pass
  through a Phillips~706 sixteen-channel discriminator. Gate and delay
  generators are used to create a veto signal which blocks the data
  during a spin reversal, allowing time for the polarization of the
  beam from the source to stabilize. The SSC determines the spin state
  of the ion source by sending TTL level signals to the ABPIS through
  fiber optic cables. The signals modulate the RF supplies of the two
  transition units. The SSC provides duplicates of these signals for
  routing information. At the end of each eight-step sequence, the
  CAMAC preset is decremented. A run is comprised of 1023 such spin
  sequences. At the end of a run, signaled by the CAMAC preset scaler
  reaching zero, the data are written to disk.
  
  Two ADC's are used for the neutron polarimeter. The first records
  the $^{4}$He recoil energy signal from the center detector. Gates
  for this ADC can come from coincidences with any of the four neutron
  detectors. Next, coincidences are formed between gates from the
  forward detectors and their corresponding PSD signals. The PSD
  coincidences for the backward detectors are generated at an earlier
  stage. Routing information is created by the ADC interface according
  to the neutron detector and the spin state. Delays are applied to
  the linear signal and the gates to allow a proper time relation to
  be established. The second ADC records the neutron time-of-flight
  signals from the four neutron detectors. The linear signals are
  fanned together using sum and invert amplifiers. The gate for this
  ADC comes from the interface for the first ADC so that the ADC's are
  triggered together and both $^{4}$He energy and neutron
  time-of-flight are recorded for each event. The events are stored in
  two-dimensional spectra with neutron time-of-flight as the $x$-axis
  and $^{4}$He recoil energy as the $y$-axis (Fig.~\ref{fig:spec}).
  There are two spectra for each of the four neutron detectors,
  corresponding to the two spin states.

\section{Data analysis}
  \label{sec:dan}
  
  The proton beam polarization is determined by observing the
  asymmetry between the number of counts from each detector,
  $\varepsilon_{p}$, for the $^{12}$C elastic peak and averaging over
  both spin states:
  \begin{equation}
    \varepsilon_{p}=\frac{1}{2}\left(
      \frac{N_{L+}-N_{R+}}{N_{L+}+N_{R+}}
      -\frac{N_{L-}-N_{R-}}{N_{L-}+N_{R-}}
    \right).
  \end{equation}
  Here, $N_{L}$ and $N_{R}$ refer to the number of counts in the left
  and right detectors and the subscripts ``$+$'' and ``$-$'' refer to
  the spin-state of the beam. The average proton beam polarization,
  $P_{p}$, is then given by
  \begin{equation}
    P_{p}=\frac{\varepsilon_{p}}{A_{y}^{p}},
  \end{equation}
  where $A_{y}^{p}$ is the analyzing power at $40^{\circ}$ for
  $^{12}$C$(\vec{p},p_{0})^{12}$C. The analyzing powers are obtained
  by fitting published analyzing power data as a function of angle
  \cite{Mos65,Ter68}. The proton polarimeter asymmetries and analyzing
  powers as well as the beam polarization are listed for each energy
  in Table~\ref{tab:ppol}.
    
  The neutron polarimeter asymmetry is calculated for each pair of
  detectors by forming the following asymmetry for the two detectors:
  \begin{eqnarray}
    \varepsilon_{n}
    & =
    & \frac{\sqrt{\frac{N_{L+}}{N_{L-}}
        \frac{N_{R-}}{N_{R+}}}-1}{\sqrt{
        \frac{N_{L+}}{N_{L-}}
        \frac{N_{R-}}{N_{R+}}}+1}, \\
    \nonumber \\
    \Delta\varepsilon_{n}
    & =
    & \frac{\sqrt{N_{L+}N_{R-}(N_{L-}+N_{R+})
        +N_{L-}N_{R+}(N_{L+}+N_{R-})}}{
      \left(\sqrt{N_{L+}N_{R-}}
        +\sqrt{N_{L-}N_{R+}}\right)^{2}}.
  \end{eqnarray}
  $N_{L}$ and $N_{R}$ refer to the number of counts in the left and
  right detectors respectively. This method of averaging reduces the
  sensitivity of the asymmetry to beam misalignments and detector
  efficiency differences. In addition, since two detectors are used,
  it is not necessary to know the incident neutron flux. The average
  neutron polarizations can be calculated from the average asymmetries
  using
  \begin{equation}
    P_{n}=\frac{\varepsilon_{n}}{A_{y}^{n}},
  \end{equation}
  where $A_{y}^{n}$ is the $^{4}$He$(\vec{n},n)^{4}$He effective
  analyzing power. The uncertainties in the analyzing powers are
  treated as systematic uncertainties. The results for each angle pair
  at each energy as well as the weighted averages over both angles are
  listed in Table~\ref{tab:npol}.
    
  Once the neutron and proton beam polarizations are known, the
  polarization-transfer coefficient is simply the ratio of the two,
  \begin{equation}
    K_{y}^{y'}=\frac{P_{n}}{P_{p}}.
  \end{equation}
  The resulting values of $K_{y}^{y'}$ for the three measurements are
  listed in Table~\ref{tab:kyy}. The proton energies listed in the
  table give the average energy and energy width after losses are
  calculated. The data are plotted in Fig.~\ref{fig:kyy}. The figure
  also includes the Los Alamos data \cite{Don71b} and the recent
  Wisconsin measurement \cite{McA94}. As can be seen in the figure,
  the data sets are in agreement.

\section{Comparison to Theory}
  \label{sec:the}
  
  The $^3$H$(p,n)^3$He reaction has a spin structure consisting of two
  spin-$\frac{1}{2}$ particles in both the entrance and exit channels.
  In this regard it is similar to nucleon-nucleon (\emph{NN})
  scattering, and the formalism developed for that problem can be used
  to describe the $^3$H$(p,n)^3$He reaction with few modifications. A
  complete description of the formalism for polarization transfer in
  the $^3$H$(p,n)^3$He reaction can be found in ref.~\cite{Hai72}.
  
  The transverse polarization transfer coefficient $K^{y'}_y(\theta)$
  is analogous to the Wolfenstein \emph{D} parameter for \emph{NN}
  scattering except that the incoming and outgoing particles are
  different. The lower index on $K^{y'}_y(\theta)$ indicates the
  component of the polarization vector for the incoming proton beam,
  while the upper index refers to the component for the outgoing
  neutron. In both instances the momentum vector of the incoming
  (outgoing) beam defines the $z$ ($z'$) axis, and the $+y$ axis is
  defined by $\vec{k}_{\rm in} \times \vec{k}_{\rm out}$. At
  $0^{\circ}$ the $y$ axis is undefined, and is taken to be in the
  vertical direction for this experiment.
  
  The \emph{M}-matrix formalism, described in detail by La France and
  Winternitz \cite{LaF80}, is well suited for the description of
  polarization observables in scattering and reaction processes.  The
  \emph{M}-matrix for the $^3$H$(p,n)^3$He reaction is a $4 \times 4$
  matrix whose elements can be written as
  \begin{equation}
    M^{s's}_{m'm} = \frac{i \sqrt{\pi}}{k} \Biggl[
    \sum_{Jll'} \sqrt{2l+1} \langle slm0 | Jm \rangle
    \langle s'l'm'(m-m') | Jm \rangle U^{J}_{l's'ls}
    Y^{m-m'}_{l'}(\theta,0) \Biggr],
  \end{equation}
  where $l$ ($l'$) is the orbital angular momentum of the incoming
  (outgoing) system, $s$ ($s'$) is the incoming (outgoing) channel
  spin, and $k$ is the wave number in the incoming channel.  The
  angular momenta are coupled according to
 \begin{eqnarray}
  J &=& (I_a + I_A) + l \nonumber \\
    &=& s + l,
 \end{eqnarray}
 where $I_a$ is the spin of the proton (neutron) and $I_A$ is the spin
 of the $^3$H ($^3$He) nucleus.  The collision matrix $U^J_{l's'ls}$
 can be written in terms of the scattering matrix as
  \begin{equation}
    U^J_{l's'ls} = -e^{i(\omega_l + \omega_{l'})} 
    S^J_{l's'ls}
  \end{equation}
  where $\omega_l$ and $\omega_{l'}$ are the modified Coulomb phase
  shifts for the incoming and outgoing waves,
  \begin{eqnarray}
    \omega_0 &=& 0 \\
  \nonumber \\
    \omega_l &=& \sum_{j=1}^{l}
    \arctan\left(\frac{\eta}{j}\right),
  \end{eqnarray}
  with the Coulomb penetration factor $\eta$ given as
  \begin{equation}
    \eta=\frac{Z_{1}Z_{2}e^{2}}{\hbar v},
  \end{equation}
  $v$ being the proton velocity.
              
  In terms of the \emph{M} matrix, $K^{y'}_y(\theta)$ is given as
  \begin{equation}
    K^{y'}_y(\theta) = \frac{{\rm Tr}(M \sigma_y^1
      M^{\dagger} \sigma_{y'}^2)} {{\rm Tr}(M M^{\dagger})}
  \end{equation}
  where $\sigma_y^1$ and $\sigma_{y'}^2$ are spin matrices for the
  incoming proton and outgoing neutrons, respectively.  At $0^{\circ}$
  all non-diagonal \emph{M}-matrix elements vanish, and
  $K^{y'}_y(0^{\circ})$ reduces to
  \begin{equation}
    K^{y'}_y(0^{\circ}) = \frac{2 {\rm Re}
   \left[M^{11 \ast}_{11}
        (M^{11}_{00} + M^{00}_{00}) \right]}
    {2|M^{11}_{11}|^2 + |M^{11}_{00}|^2
   + |M^{00}_{00}|^2}.
  \label{eq:simple}
  \end{equation}
    
  Since $K^{y'}_y(0^{\circ})$ is sensitive to relatively few matrix
  elements, its behavior can be analyzed in terms of the excited
  states of the $^4$He compound nucleus. In Fig.~\ref{fig:kyy} we
  compare the present and previous results for the $0^{\circ}$
  transfer coefficient to predictions based on a charge-independent
  \emph{R}-matrix analysis of $A=4$ scattering and reaction data
  \cite{Til92}. Neither the present nor previous results were included
  in the \emph{R}-matrix analysis.
    
  As indicated by Eq.~\ref{eq:simple}, $K^{y'}_y(0^{\circ})$ results
  from the spin-triplet matrix element $M^{11}_{11}$ interfering with
  either $M^{11}_{00}$ or $M^{00}_{00}$.  For proton energies below
  about 2~MeV, the $^3$H$(p,n)^3$He reaction is dominated by a pair of
  isospin-0 resonances in the $^4$He compound nucleus.  The spins and
  parities of these resonances are $0^+$ and $0^-$.  Their resonance
  energies, according to the \emph{R}-matrix analysis, are 20.21 and
  21.01~MeV above the $^4$He ground state, respectively. Neither of
  these resonances contributes to the $M^{11}_{11}$ matrix element,
  and $K^{y'}_y$ falls rapidly below 2~MeV\@.
    
  Above 2~MeV the $M^{11}_{11}$ matrix element rises rapidly and has a
  broad peak at $E_p=3.7$~MeV\@.  The $M^{11}_{00}$ matrix element
  displays similar behavior, which is due to the emergence of a $2^-$
  isodoublet at 21.84 and 23.33~MeV\@.  The $2^-$ resonances
  contribute to both $M^{11}_{11}$ and $M^{11}_{00}$, and it is the
  interference of these resonances with one other that is responsible
  for the sharp increase in $K^{y'}_y$ above 2~MeV\@.
    
  In addition to the low-energy $0^-$ and the $2-$ states, there are
  four additional \emph{P}-wave resonances that are primarily
  \emph{p}-$^3$H (and \emph{n}-$^3$He) in character.  A second $0^-$
  state, with isospin $T=1$, exists at 25.28~MeV and produces negative
  values of $K^{y'}_y$.  Removing this state from the analysis results
  in a systematic increase in $K^{y'}_y$ at all energies.  A $1^-$
  isodoublet, $T=1$ at 23.65 and $T=0$ at 24.25~MeV, has little effect
  on $K^{y'}_y$.  Both resonances are essentially pure spin-triplet
  states, and the $^3\!P_1$ partial wave cannot contribute to the
  $M^{11}_{00}$ matrix element.  While this partial wave does
  contribute to $M^{11}_{11}$, particularly around $E_p=4$~MeV,
  $M^{11}_{11}$ is dominated by the $^3\!P_2$ partial wave from the
  $2^-$ resonances.  A third $1^-$, state ($T=0$) exists at 25.95~MeV,
  and is primarily spin-singlet.  This state has a profound effect on
  $K^{y'}_y$ values above 2~MeV\@.  Upon removal of the $^1\!P_1$
  partial wave, $K^{y'}_y$ still exhibits a sharp rise due to the
  $2^-$ states, but falls to nearly zero at 4~MeV\@.  The interference
  of the $1^-$ state at 25.95~MeV with the $2^-$ states at 21.84 and
  23.33~MeV generates large values of $K^{y'}_y$ above 4~MeV.
    
  While the \emph{R}-matrix analysis is able to describe the
  qualitative trend of the experimental behavior, it predicts values
  that are substantially lower at energies above 5~MeV\@. We find that
  changing the sign of of the \emph{R}-matrix $^3\!F_3$ phase shift
  from negative to positive, improves the agreement.  This has
  comparatively minor effects on other $^{3}$H$(p,n)^{3}$He
  observables, notably the analyzing power $A_y$. It is likely that
  minor changes can be made to the other phases to compensate for the
  $^3\!F_3$ phase shift.

\section{Summary}
  \label{sec:sum}
  
  We have made measurements of the polarization-transfer coefficient
  $K_{y}^{y'}$ at $0^{\circ}$ for the reaction
  $^{3}$H$(\vec{p},n)^{3}$He at three neutron energies in the range
  1.94 to 5.81~MeV\@. This reaction is used at TUNL and other
  facilities for producing polarized neutron beams at few-MeV
  energies.  Our measurements confirm earlier results from Los Alamos
  National Laboratory and extend to lower energies.  Comparing
  \emph{R}-matrix predictions to the experimental data reveals a
  discrepancy which we attribute to the sign of the $^{3}\!F_{3}$
  phase-shift parameter.  Changing the sign of this parameter from
  negative to positive dramatically improves the agreement between
  theory and experiment, while having little effect on other
  observables.

\section{Acknowledgments}
  \label{sec:ack}
  
  This research was supported in part by the U.S. Department of
  Energy, Office of High Energy and Nuclear Physics, under Grants No.\ 
  DEFG05-91-ER40619 and DEFG05-88-ER40441.

\begin{table}
  \begin{tabular}{dddr@{${}\pm{}$}lr@{${}\pm{}$}l}
    \hline
    $E_{n}$ (MeV)
    & $E_{p}$ (MeV)
    & $\varepsilon_{p}$
    & \multicolumn{2}{c}{$A_{y}^{p}$}
    & \multicolumn{2}{c}{$P_{p}$}
    \\
    \hline
    1.94
    & 3.02
    & 0.401
    & $-$0.851
    & 0.009
    & $-$0.471
    & 0.05
    \\
    5.21
    & 6.18
    & 0.395
    & $-$0.851
    & 0.009
    & $-$0.464
    & 0.005
    \\
    5.81
    & 6.77
    & $-$0.316
    & $-$0.520
    & 0.009
    & 0.608
    & 0.011
    \\
    \hline
  \end{tabular}
  \caption{Proton polarimeter data. Left-right scattering
   asymmetry, analyzing power, and proton polarization
   with systematic uncertainties are listed for each
   energy. The statistical uncertainties in the
   asymmetries are negligible.}
 \label{tab:ppol}
\end{table}

\begin{table}
  \begin{tabular}{drr@{${}\pm{}$}lr@{${}\pm{}$}lr@{${}\pm{}$}l
      @{${}\pm{}$}l}
    \hline
    $E_{n}$ (MeV)
    & \multicolumn{1}{c}{$\theta$}
    & \multicolumn{2}{c}{$\varepsilon_{n}$}
    & \multicolumn{2}{c}{$A_{y}^{n}$}
    & \multicolumn{3}{c}{$P_{n}$}
    \\
    \hline
    1.94
    & $107^{\circ}$
    & $-$0.2362
    & 0.0122
    & 0.764
    & 0.010
    & $-$0.309
    & 0.016
    & 0.004
    \\
    \\
   5.21
   & $50^{\circ}$
   & 0.2405
   & 0.0166
   & $-$0.629
   & 0.010
   & $-$0.382
   & 0.026
   & 0.006
   \\
   & $121^{\circ}$
   & $-$0.3163
   & 0.0277
   & 0.919
   & 0.015
   & $-$0.344
   & 0.030
   & 0.006 \\
   & \multicolumn{5}{c}{Average}
   & $-$0.366
   & 0.020
   & 0.006
   \\
   \\
   5.81 &  $51^{\circ}$
   & $-$0.3045
   & 0.0300
   & $-$0.632
   & 0.010
   & 0.482
   & 0.048
   & 0.008
   \\
   & $121^{\circ}$
   & 0.4594
   & 0.0560
   & 0.916
   & 0.015
   & 0.502
   & 0.061
   & 0.008
   \\
   & \multicolumn{5}{c}{Average}
   & 0.490
   & 0.038
   & 0.008
   \\
   \hline
 \end{tabular}
 \caption{Neutron polarimeter data. Left-right scattering
   asymmetry, analyzing power, and neutron polarization
   with statistical and systematic uncertainties are
   listed for each energy and detector pair. The
   uncertainties in the analyzing powers are treated as
   systematic. The measured polarization averaged over
   both detector pairs is also given.}
 \label{tab:npol}
\end{table}

\begin{table}
  \begin{tabular}{dr@{${}\pm{}$}lr@{${}\pm{}$}l@{${}\pm{}$}l}
    \hline
    $E_{\mathrm{n}}$ (MeV)
    & \multicolumn{2}{c}{$E_{p}$ (MeV)}
    & \multicolumn{3}{c}{$K_{y}^{y'}$} \\
    \hline
    1.94
    & 3.02
    & 0.11
    & 0.656
    & 0.034
    & 0.011
    \\
    5.21
    & 6.18
    & 0.08
    & 0.789
    & 0.043
    & 0.015
    \\
    5.81
    & 6.77
    & 0.07
    & 0.806
    & 0.063
    & 0.020
    \\
    \hline
  \end{tabular}
  \caption{The measured values of $K_{y}^{y'}$ with statistical
   and systematic uncertainties. The uncertainty in
   $E_{p}$ reflects the energy width of the proton beam
   due to losses.}
 \label{tab:kyy}
\end{table}

\begin{figure}
  \caption{The proton polarimeter showing the carbon foil and the
    two silicon detectors. Also shown are the tantalum collimators.}
  \label{fig:ppol}
\end{figure}

\begin{figure}
  \caption{The neutron polarimeter showing the $^{4}$He gas cell
    and the forward and backward pairs of neutron detectors.}
  \label{fig:npol}
\end{figure}

\begin{figure}
  \caption{The high-pressure $^{4}$He gas cell.}
  \label{fig:hecl}
\end{figure}

\begin{figure}
  \caption{The gas filling system for the $^{4}$He cell.}
  \label{fig:hefl}
\end{figure}

\begin{figure}
  \caption{A representative two-dimensional spectrum from the
    neutron polarimeter. The $x$-axis is neutron time-of-flight with
    time increasing to the left, while the $y$-axis is $^{4}$He recoil
    pulse height. The number of counts is shown in the $z$~direction.}
  \label{fig:spec}
\end{figure}

\begin{figure}
  \caption{The experimentally determined values of $K_{y}^{y'}$
    from the present work (triangles), the Wisconsin measurement
    \protect\cite{McA94}, and the previous Los Alamos data (circles)
    \protect\cite{Don71b}. The error bars indicate the total
    uncertainty obtained by adding the statistical and systematic
    uncertainties in quadrature.  The lines show the \emph{R}-matrix
    predictions with (solid) and without (dashed) modification of the
    $^{3}\!F_{3}$ phase-shift parameter, as explained in the text.}
  \label{fig:kyy}
\end{figure}

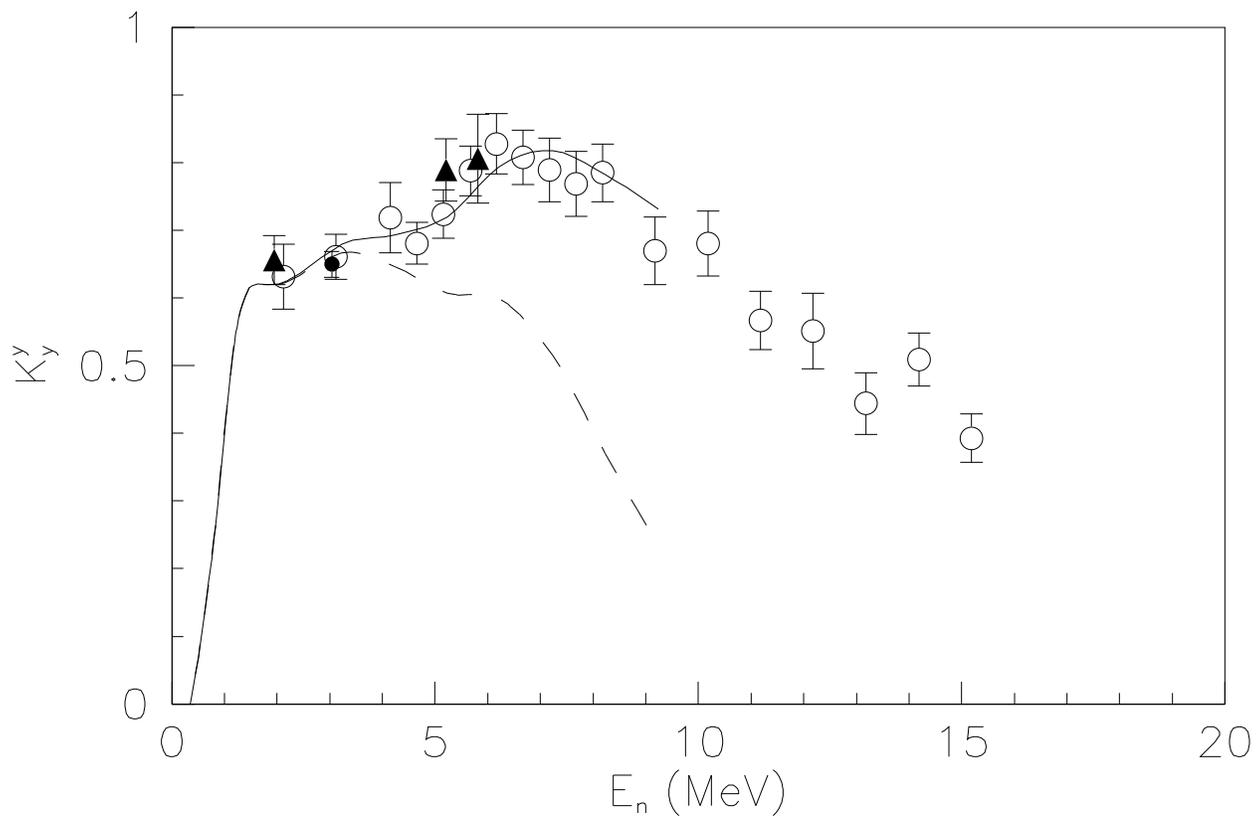
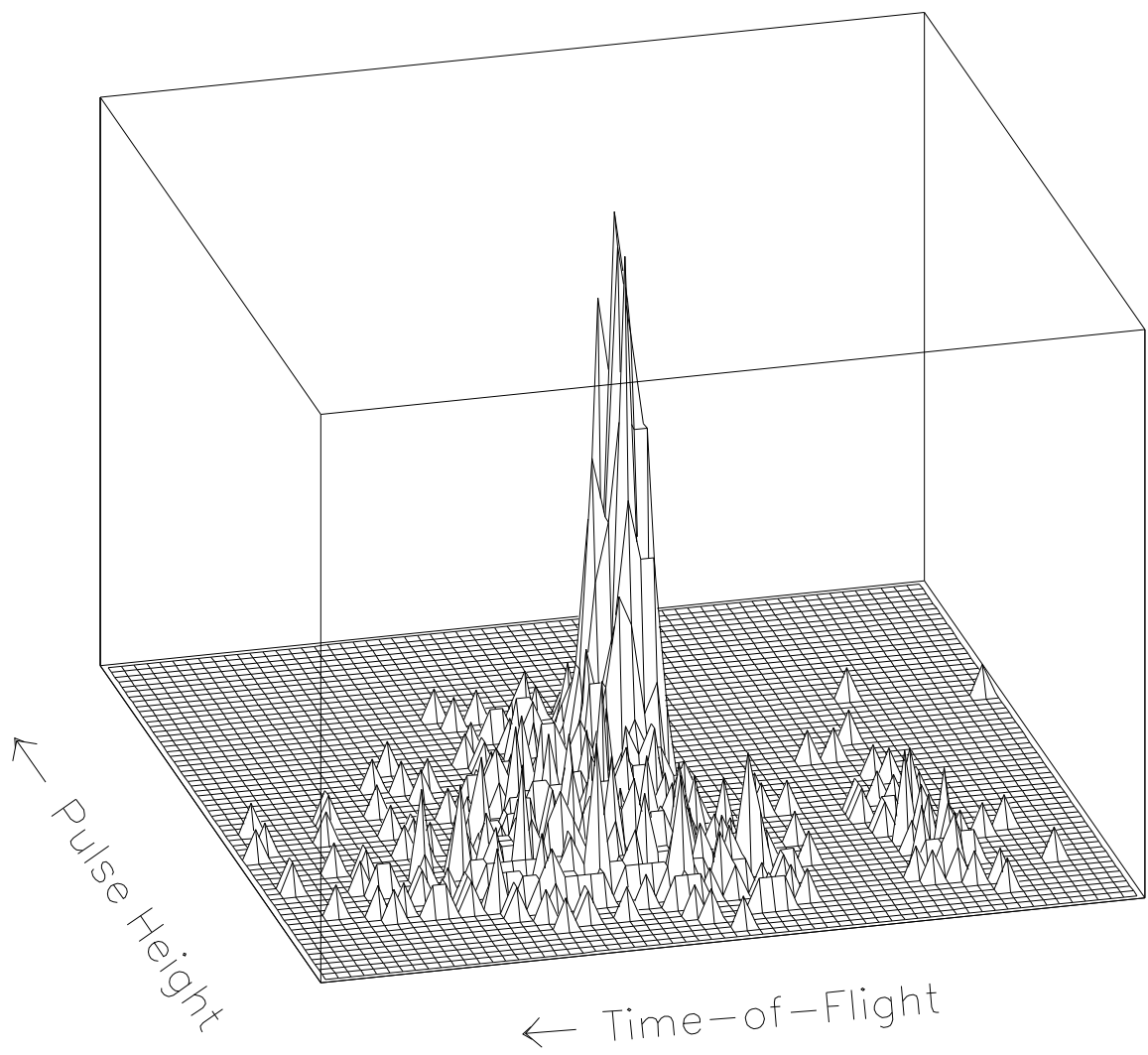
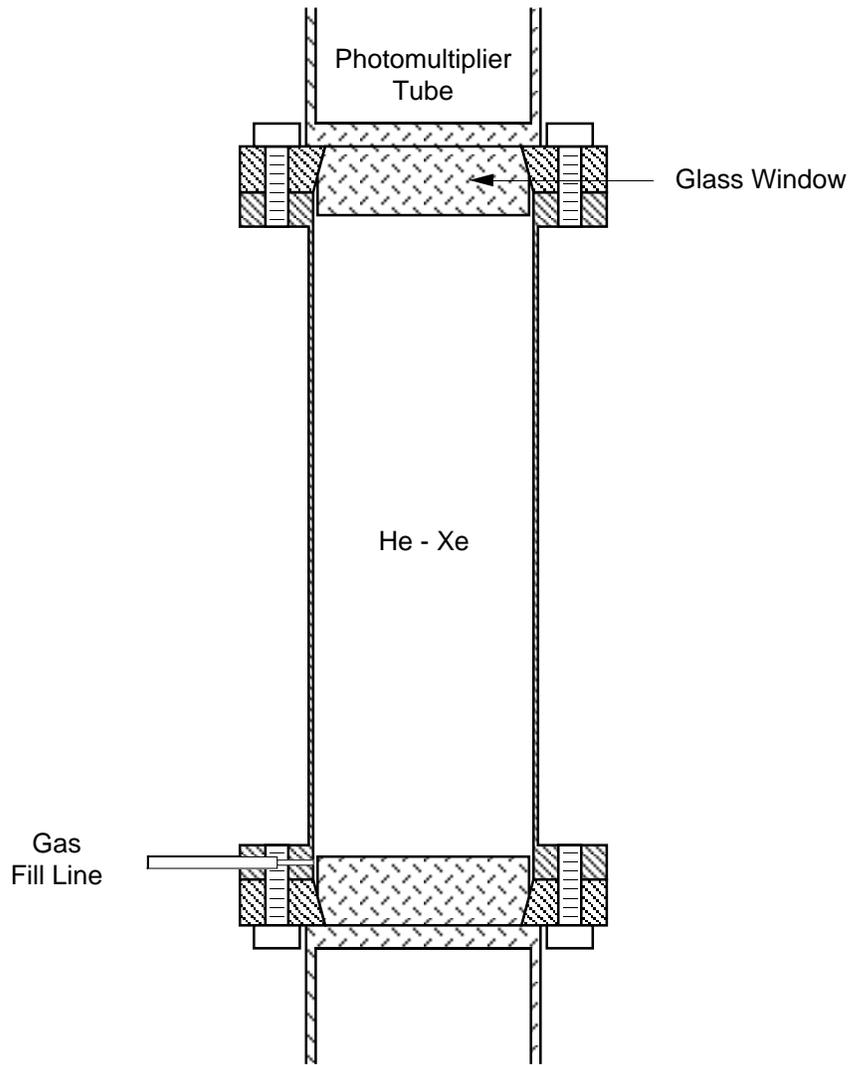
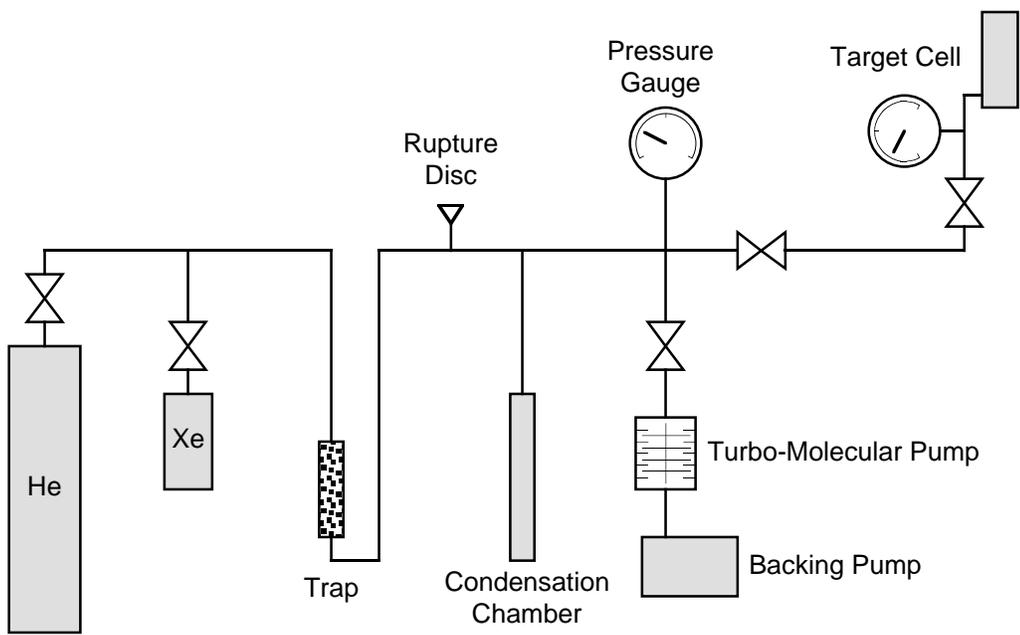
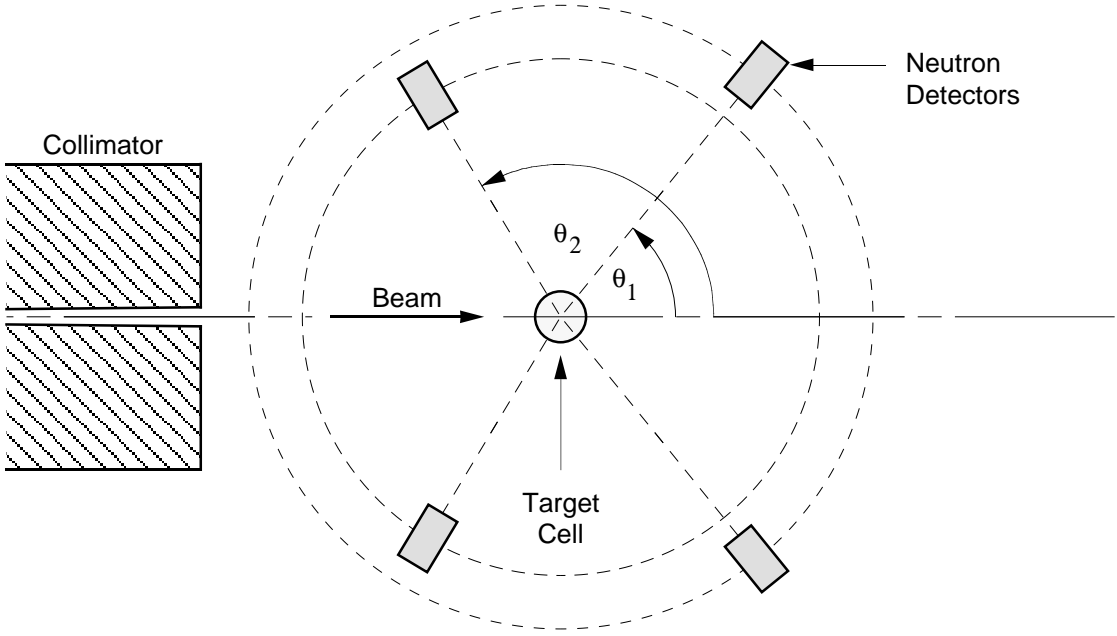
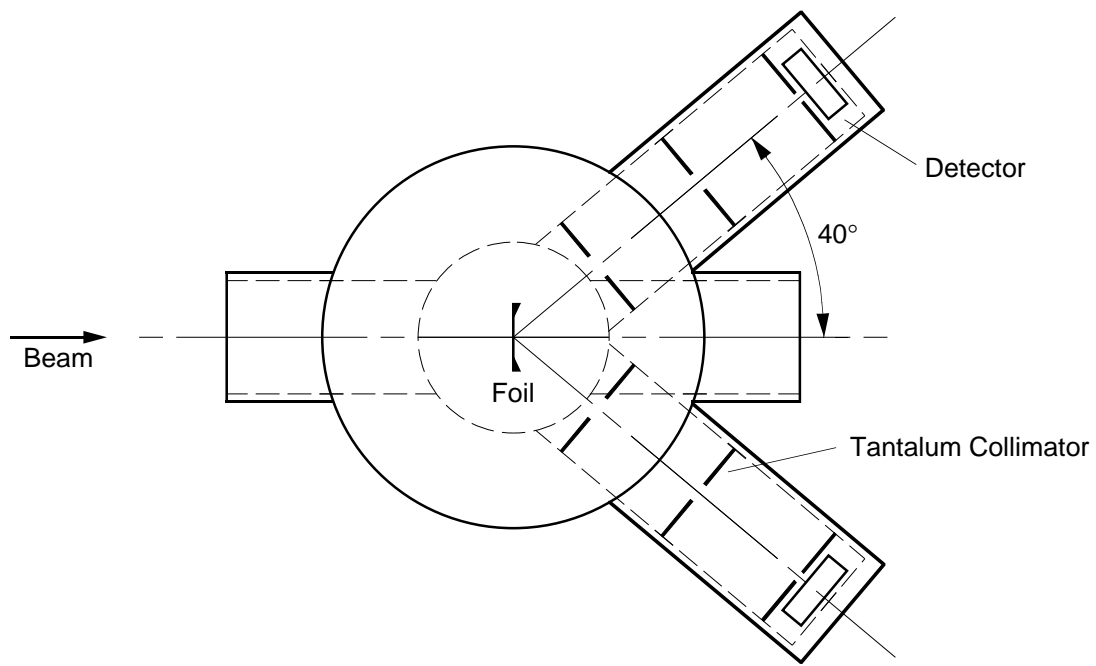
\end{document}